\documentclass[11pt, epsfig]{article}
\usepackage{epsfig, amsmath, amssymb, amsthm, times}
\usepackage{setspace}

\def\theequation{\arabic{section}.\arabic{equation}}
\renewcommand{\theequation}{\thesection.\arabic{equation}}

\newtheorem {thm}{Theorem}

\newtheorem {cor}[thm]{Corollary}

\theoremstyle{defintion}
\newtheorem {df}[thm]{Definition}

\theoremstyle{remark}

\theoremstyle{example}

\def\pf{{\it Proof.\;}}

\def\var{{\rm var~}}
\def\ovar{\overline{\rm var}~}
\def\cov{{\rm cov~}}

\def\P{{\mathbb P~}}
\def\Re{{\mathbb R}}

\def\Co{{\mathbb C}}

\def\lbl{\label}
\def\be{\begin{equation}}
\def\ee{\end{equation}}
\def\qed{\square}
\def\one{\mathbf{1_N}}
\def\tr{{\bf Tr~}}
\def\cl{{\mathit{l}}}

\title{Synchronization of coupled stochastic limit cycle oscillators}
\author{Georgi S. Medvedev
\thanks{
Department of Mathematics, Drexel University, 3141 Chestnut Street,
Philadelphia, PA 19104, {\tt medvedev@drexel.edu}, 
ph.: 1-215-895-6612, fax: 1-215-895-1582} }
\date{\today}
\begin{document}
\maketitle
\begin{abstract}
For a class of coupled limit cycle oscillators,
we give a condition on a linear coupling operator that is
necessary and sufficient for exponential stability of the synchronous
solution. We show that with certain modifications our method of analysis
applies to networks with partial, time-dependent, and nonlinear
coupling schemes, as well as to ensembles of local systems
with nonperiodic attractors. We also study robustness of synchrony to noise. 
To this end,  we analytically estimate the degree of coherence of
the network oscillations in the presence of noise.
Our estimate of coherence highlights the main ingredients of stochastic
stability of the synchronous regime. In particular, it quantifies
the contribution of the network topology. The estimate of coherence
for the randomly perturbed network can be used 
as means for analytic inference of degree of stability of the 
synchronous solution of the unperturbed deterministic network.
Furthermore, we show that in large networks, the effects of noise on 
the dynamics of each oscillator can be effectively
controlled by varying the strength of coupling, which provides a powerful
mechanism of denoising. This suggests that the organization of oscillators
in a coupled network may play an important role in maintaining 
robust oscillations in random environment. 
The analysis is complemented with the results of numerical
simulations of a neuronal network. 
\end{abstract} 

\section{Introduction}
\setcounter{equation}{0}
Consider a dynamical system forced by small noise:
\be\lbl{2.1}
\mathsf{\dot x_t=f(x_t)+\sigma P(t) \dot w_t},\;
\mathsf{x}:\Re^1\to\Re^n,
\ee
where function $\mathsf{f}:\Re^n\to\Re^n$ is continuous together with 
partial derivatives up to the second order, $\mathsf{P:} \Re^1\to\Re^n$ is 
a bounded continuous function of time, $\mathsf{\dot w_t}$ is
a $n-$dimensional white noise process, and small
$\sigma>0$ is the noise intensity. We call
(\ref{2.1}) a local system and denote (\ref{2.1}$)_0$ the underlying
deterministic system (\ref{2.1}) with $\sigma=0$. For $\sigma>0$, 
stochastic differential equation (\ref{2.1}) is understood in the 
Ito's sense \cite{KS}.
Many models of (bio)physical 
phenomena are formulated as the coupled networks of $N$ identical 
local systems (\ref{2.1}):
\be\lbl{1.2}
\dot x_t=f(x_t)+gDx_t +\sigma P(t)\dot w_t,
\ee
where $x=\left(\mathsf{x^{(1)},x^{(2)},\dots,x^{(N)}}\right)\in\Re^{Nn}$,
$f(x)=\left(\mathsf{f(x^{(1)}),f(x^{(2)}),\dots,f(x^{(N)})}\right)\in\Re^{Nn}$
and\\
$P(t)=\mathbf{I_N}\otimes\mathsf{P(t)}$, and
 $w_t=\left(\mathsf{w_t^{(1)},w_t^{(2)},\dots,w_t^{(N)}}\right)$, $\mathsf{w_t^{(i)}}$ 
are independent copies of $n-$dimensional Brownian motion.  
The coupling is implemented by a linear operator
$D:\Re^{Nn}\rightarrow\Re^{Nn}$ and $g\ge 0$ is interpreted as the strength
of coupling. $D$ may depend on time. 
Our only assumption on $D$ is that it leaves the diagonal
invariant, i.e., if $\mathsf{x=\xi(t)}$ solves (\ref{2.1}$)_0$ then
$x=\xi(t):=\mathbf{1_N}\otimes\mathsf{\xi(t)}$ solves (\ref{1.2}$)_0$.
Here, $\mathbf{1_N}=(1,1,\dots,1)^T\in\Re^N$ and $\otimes$ denotes the Kronecker
product, so that $\xi(t)=(\mathsf{\xi(t),\xi(t),\dots,\xi(t)})$
is a solution of the coupled system. 
Such solutions, when asymptotically stable, feature  synchronization, an important mode
of collective behavior. Suppose $\mathsf{x=\xi(t)}$
is an asymptotically stable solution of the local system (\ref{2.1}$)_0$.
Under what conditions on the coupling operator $D$, $x=\mathbf{1_N}\otimes\mathsf{\xi(t)}$
is an asymptotically stable solution of (\ref{1.2}$)_0$? What information about
$D$ is important for synchronization properties of the coupled system?  What determines 
the rate of attraction of the coupled limit cycle and its robustness to noise? 
Clearly, the answers to these questions
depend in a nontrivial fashion on the properties of the local systems,
network topology, and the type of interactions between local systems. 

Many different approaches have
been proposed to study synchronization: asymptotic analysis \cite{avr, bh84}, 
phase reduction \cite{ke88, ku75}, 
constructing Lyapunov functions \cite{ste} and estimating Lyapunov exponents \cite{pc98},
 invariant manifold theory \cite{hale97, jo}, 
and  graph-theoretic techniques \cite{be}. Above we selected just a few representative
studies illustrating these approaches. For more background 
and more complete bibliography, we refer an interested reader to recent monographs 
\cite{pr01, mmp, stro}. For weakly coupled networks, i.e., when $0<g\ll 1$, there 
are effective perturbation techniques for studying synchronization 
such as asymptotic approximation of the Poincare map
and the method of averaging. These ideas underlie widely used method of phase response
curves and Kuramoto's phase reduction \cite{ku75, ke88}.
For a review of techniques available for studying synchronization
in weakly coupled networks we refer to Chapter 10 in 
\cite{izhikevich} and references therein. When the coupling is moderate 
$g=O(1)$ or strong, these methods do not apply. Moreover, the mechanisms
for synchronization in weakly and strongly coupled networks are different.
In the case of strong coupling, a prevalent approach for studying
synchronization is to use the properties of specific 
(albeit important in applications) coupling schemes such as global all-to-all
coupling (see, e.g., \cite{coombes}), local nearest-neighbor coupling,
or more generally schemes resulting from discretization of Laplace
operator \cite{avr, hale97}. For these network topologies, one can use
explicit information about the spectra of the coupling matrices; in addition,
the former scheme has strong symmetry properties that can be used
in understanding network dynamics. Networks with general coupling 
operators have been studied by Pecora and Caroll \cite{pc98}
and by V.~Belykh, I.~Belykh, and Hasler \cite{be, bbh06}.
The master stability function, constructed in \cite{pc98}, uses
spectral properties of a given coupling operator to determine whether
the synchronous solution is stable. Practical implementation of this
method relies on numerical computation of matrix eigenvalues.
Analytical sufficient conditions for synchronization derived in \cite{be, bbh06}
use graph theoretic interpretation of the coupling operator
to construct Lyapunov functions controlling the growth of perturbations 
of the synchronous solution. In this Letter, we look for an analytic
or rather algebraic description of coupling operators that endow 
synchronous solutions with exponential stability. 
 We define a class of matrices,
dissipative matrices (cf. Definition~\ref{df.1}), and show that these 
matrices generate exponentially stable synchronous 
solutions once the coupling strength exceeds a certain value.
A random dissipative matrix shown in Section~\ref{numerical} 
suggests that many  dissipative
matrices do not fall into the class of coupling operators analyzed
in \cite{be, bbh06}. Therefore, by identifying dissipative matrices
we have substantially extended existing knowledge of linear coupling
operators that enforce synchrony in coupled networks.
Surprisingly, dissipative matrices admit an explicit algebraic
characterization: Theorem~\ref{thm.ql} relates all dissipative matrices to 
a discrete Laplacian, justifying common interpretation of the Laplacian  
as a prototypical diffusive coupling. To highlight the main ingredients
of our treatment of synchronization, in Section~\ref{analysis} 
we present the analysis in the
simplest meaningful setting when the coupling is linear and
stationary. In Section~3, we discuss how to apply our method
to problems with time-dependent and nonlinear coupling operators,
and ,importantly, to networks composed of local systems with
nonperiodic attractors.

Adequate description of many physical phenomena requires including
stochastic terms into differential equation models. In the context
of synchronization this leads to an important question of robustness
of synchrony to noise. This is the second problem investigated in this
Letter. We analytically
estimate the coherence of the coupled oscillations in the presence 
of noise. The estimate
is tight. It reflects the main ingredients of robustness of 
synchronous oscillations to noise.
In particular, it quantifies the contribution of the network topology to the stability
of the synchronous solution.
As a related result, we show that in large networks the effects 
of noise on oscillations can be reduced substantially 
by increasing the strength of coupling. For networks of simpler
elements, so-called integrate and fire neurons, the denoising property
was shown in \cite{M09}. The present Letter extends the result
of \cite{M09} to systems of coupled limit cycle oscillators.
Finally, in Section~4 we illustrate our findings with a discussion
of the dynamics of a concrete biophysical model, an ensemble
of neural oscillators.

\section{The analysis}\lbl{analysis}
\setcounter{equation}{0}

We start by specifying the structure of the coupling operator. 
We call coupling operator $D$ {\it separable} if 
\be\lbl{sep}
D=\mathbf{D}\otimes\mathsf{L},\; \mathbf{D}\in\Re^{N\times N}, 
\mathsf{L}\in\Re^{n\times n}.
\ee
Matrices $\mathbf{D}$ and
$\mathsf{L}$ play distinct roles in the network organization. 
$\mathbf{D}$ reflects the global architecture: 
what local system is connected
to what. $\mathsf{L}$ specifies how the coupling is organized on the level
of a local system: roughly, what local variables are engaged in coupling.
The separable structure of the coupling is important. 
It translates naturally to
the stability analysis of the synchronous state.
The condition that the diagonal is invariant for separable
coupling translates to
$
\mathbf{1_N}\in\ker~\mathbf{D}.
$
Moreover, if the network is connected then $\ker~\mathbf{D}$
is one-dimensional. Thus, we are led to the following condition
\be\lbl{kerD}
\mathbf{D}\in\mathcal{K}=\left\{\mathbf{M}\in\Re^{N\times N}:~
\ker~\mathbf{M}=\mbox{Span}~\{\mathbf{1_N} \}\right\}.
\ee 
Further, we assume that $\mathsf{L}$ is symmetric  semipositive
definite, i.e., $\mathsf{L^T=L}$ and $\mathsf{x^TLx\ge 0}$
$\mathsf{\forall x\in\Re^n}$. 
$\mathbf{D}$ is not assumed symmetric. For simplicity, we keep 
$\mathbf{D}$ and $\mathsf{L}$ constant. In Section~\ref{generalizations}, 
we explain our results for time-dependent and nonlinear coupling.

If $\mathsf{L}$ is positive definite,
we say that the coupling is full (rank), otherwise we call it
partial (rank). The distinction between the full and partial 
coupling is important for synchronization properties of (\ref{1.2}).
The following examples illustrate how full and partial coupling 
arise in physical models. Let
\be\lbl{all}
\mathbf{D_1}=\left(\begin{array}{ccccc}
-N+1 &1 & 1& \dots &1 \\
1 & -N+1& 1 & \dots& 1\\
\dots&\dots&\dots&\dots&\dots\\
1 & 1 & 1& \dots &-N+1
\end{array}
\right),
\ee
$\mathsf{I}$ be the $n\times n$ identity matrix, and 
$\mathsf{I^\prime}=\mbox{diag}(1,0,0,\dots,0)\in \Re^{n\times n}$. 
$D:=\mathbf{D_1}\otimes\mathsf{I}$ in (\ref{1.2}) implements
all-to-all coupling through all variables, while  
$D:=\mathbf{D_1}\otimes\mathsf{I^\prime}$
engages only the first variables of the local systems.
The former is an example of the full coupling, whereas the latter is
that of the partial coupling. The stability analysis of partially coupled systems
has to deal with the degeneracy of the coupling matrix $D$ (due to the zero 
eigenvalues of $\mathsf{L}$). A reader wishing to gain better physical intuition
for coupled system (\ref{1.2}) and (\ref{sep}) before embarking on analysis,
is referred to the discussion of a compartmental model of a neuron in 
Section~\ref{numerical}.

To study synchronization in (\ref{1.2})-(\ref{kerD}), we derive the equation
for the phase variables of the coupled system. To set up the notation,
we review phase reduction for a single oscillator.
Let $\mathsf{x=\xi(t)}$ denote a periodic solution of (\ref{2.1}$)_0$ with the
least period $1$.
By $\mathsf{\mathcal{O}=\{x=\xi(t),\; t\in S^1=\Re/\mathbb{Z}\}}$ 
we denote the corresponding 
orbit. 
Along $\mathcal{O}$ we introduce an orthonormal moving coordinate frame 
(cf. \cite{hale_odes}):
\be\lbl{7.1}
\mathsf{ \{ v(\theta), z_1(\theta), z_2(\theta),\dots, z_{n-1}(\theta)\},\;
\theta\in S^1},
\ee
where the first vector is a unit vector moving along $\mathcal{O}$,
$\mathsf{v(\theta)=\dot\xi(\theta)\left|\dot\xi(\theta)\right|^{-1}}$.
The change of variables 
\be\lbl{7.2}
\mathsf{
x=\xi(\theta)+Z(\theta)\rho,\; 
Z(\theta)=\mbox{col}(z_1(\theta),\dots, z_{n-1}(\theta)).
}
\ee
defines a smooth transformation $\mathsf{x\mapsto (\theta,\rho)\in S^1\times \Re^{n-1}}$ 
in a sufficiently small neighborhood of $\mathcal{O}$ 
(cf. Theorem VI.1.1 in \cite{hale_odes}). Using Ito's formula,
in the vicinity of the periodic orbit, we rewrite (\ref{2.1})
in terms of $\theta$ and $\rho$ (cf. (\ref{7.2}))
and project the resultant equation onto the subspaces
spanned by $\mathsf{v(\theta)}$ and 
$\{\mathsf{z_1(\theta),\dots, z_{n-1}(\theta)}\}$
to obtain
\begin{eqnarray}\lbl{7.3}
\mathsf{\dot\theta_t}&=& \mathsf{1+\sigma h_1(\theta_t) P(t)\dot w_t +\dots},\\
\lbl{7.4}
\mathsf{\dot\rho_t} &=&\mathsf{ A(\theta_t)\rho_t+\sigma h_2(\theta_t)P(t)\dot w_t +
\dots},
\end{eqnarray}
where
\be\lbl{7.5}
\mathsf{A(\theta)=
Z(\theta)^T\left[ {-\partial Z(\theta)\over \partial\theta}+ 
Df\left(\xi(\theta)\right)Z(\theta)
\right]},
\ee
$$
\mathsf{
h_1(\theta)=v^T(\theta)|\dot\xi(\theta))|^{-1},\;
h_2(\theta)=Z^T(\theta).}
$$
The detailed derivation of the system of equations near a periodic orbit 
of a deterministic system in the moving coordinates can be found in the 
proof of Theorem VI.1.2 in \cite{hale_odes}. The treatment of the stochastic
case requires Ito's formula, which does not affect the leading 
order terms written out in (\ref{7.3}) and (\ref{7.4}) 
(see the proof of Lemma~4.1 in \cite{HM} for details).
The solution of an initial value problem for (\ref{7.3}) and (\ref{7.4})
yields a real-valued function $\mathsf{\theta_t}$. The phase of the
oscillations is given by $(\mathsf{\theta_t}~\mbox{mod}~1)\in\mathsf{S^1}$. 
It will be more convenient to work with $\mathsf{\theta_t}$, 
which we will call a phase variable,
rather than with it's projection on $\mathsf{S^1}$.
 
Assume that the eigenvalues of $A^s(\theta)=A(\theta)+A^T(\theta)$, 
$\lambda_i(\theta)$, $i=1,2,\dots,n-1$, are negative
\be\lbl{7.5a}
\max_{\theta\in S^1} \lambda_i(\theta)\le -\bar\lambda <0.
\ee
By applying the phase reduction to each oscillator in the network,
in complete analogy  to (\ref{7.3}), we derive the phase equations
for the coupled system
$$
\mathsf{\dot\theta^{(i)}_t}=\mathsf{1+
\sigma h_1(\theta_t^{(i)})P(t)\dot w^{(i)}_t}
$$
\be\lbl{7.6}
+g\mathsf{{v^T(\theta^{(i)}_t)\over |\dot\xi(\theta^{(i)})|}\sum_{i\ne j} \mathbf{d_{ij}} L 
 \left(\xi(\theta_t^{(j)})-\xi(\theta^{(i)}_t)\right)+\dots
},
\ee
where $\mathbf{d_{ij}}$ denote the entries of $\mathbf{D}$ (cf. (\ref{sep})).
The expression for the coupling terms in (\ref{7.6}) simplifies to  
$$
\mathsf{ 
 {v^T(\theta^{(i)}) \over |\dot\xi(\theta^{(i)})|} \mathsf{L}
\left[ \xi (\theta^{(j)})-\xi (\theta^{(i)}) \right]
}
=
\mathsf{
{\dot\xi (\theta^{(i)})^T \mathsf{L} \dot\xi (\theta^{(i)})
\over \left| \dot\xi (\theta^{(i)})\right|^2} \times
}
$$
\be\lbl{7.8}
\mathsf{
\times\left(\theta^{(j)}-\theta^{(i)}\right)+\dots
}
\ee
Here and below, we ignore quadratic  terms
$O\left((\theta^{(i)}-\theta^{(j)})^2\right)$. 
By plugging (\ref{7.8}) in (\ref{7.6}), we arrive at the following system
of equations
\be\lbl{7.9}
\dot\theta_t =\one +gV_N(\theta_t)\mathbf{D}\theta_t+\sigma H_1(\theta_t)P(t)\dot w_t
+\dots ,
\ee
where 
$$
V_N(\theta)=\mbox{diag}
\left( \cl(\mathsf{\theta^{(1)}}),\cl(\mathsf{\theta^{(2)}}),
\dots,\cl(\mathsf{\theta^{(N)}})\right),
H_1(\theta)=\mbox{diag}
\left( \mathsf{h_1(\theta^{(1)}),\dots,h_1(\theta^{(N)})}\right),
$$
and
\be\lbl{ltheta}
\cl(\mathsf{\theta}):= 
\mathsf{\left|\dot\xi(\theta)\right|^{-2}  
\dot\xi(\theta)^T\mathsf{L}\dot\xi(\theta)
=v(\theta)^TLv(\theta). 
}
\ee
Next, we derive the system for the vector of the phase differences
\be\lbl{phi}
\phi=\mathbf{S}\theta=\left(
\mathsf{\phi^{(1)},\dots,\phi^{(N-1)}}\right),\; 
\mathsf{\phi^{(i)}=\theta^{(i+1)}-\theta^{(i)}},
\ee
where $(N-1)\times N$ matrix $\mathbf{S}$ is defined by
\be\lbl{S}
\mathbf{S}=
\left(\begin{array}{cccccc}
-1 & 1 & 0& \dots &0& 0 \\
0 & -1& 1 & \dots& 0& 0\\
\dots&\dots&\dots&\dots&\dots&\dots\\
0 &0 & 0& \dots &-1 & 1
\end{array}
\right).
\ee
Multiply both sides of (\ref{7.9}) by $\mathbf{S}$ and note that
\be\lbl{7.10}
\mathbf{S}\one=0\quad\mbox{and}\quad 
\mathbf{SV_N}\theta=\mathbf{V_{N-1}S}\theta+O(|\phi|),
\ee 
where $\phi$ is defined in (\ref{phi}).
From (\ref{7.9}) and (\ref{7.10}) we have
\be\lbl{7.11}
\dot\phi_t =g V_{N-1}(\theta_t)\mathbf{\hat D} \phi_t+ 
\sigma \mathbf{S}H_1(\theta_t)P(t)\dot w_t + 
\dots,
\ee
where $\mathbf{\hat D}$ is defined by $\mathbf{SD}=\mathbf{\hat DS}$.
For $\mathbf{D}\in\mathcal{K}$, $\mathbf{\hat D}$ is well-defined
(cf. Appendix \cite{M09}). 
In the derivation of (\ref{7.11}), we treated 
terms $\sim |\phi|\mathbf{\hat D}\phi=O(|\phi|^2)$ inherited from 
(\ref{7.10}) as higher order terms.
Since $\theta^{(i)}=\theta^{(1)}+O(\left|\phi\right|)$
(cf. (\ref{phi})),
\begin{eqnarray*}
V_{N-1}(\theta)&=&\cl(\theta^{(1)})\mathbf{I_{N-1}}+O(\left|\phi\right|),\\
H_1(\theta)&=&\mathbf{I_N}\otimes h_1(\theta^{(1)})+O(\left|\phi\right|),
\end{eqnarray*}
and (\ref{7.11}) is reduced to
\begin{eqnarray}\nonumber
\dot\phi_t &=& g \cl(\theta^{(1)}_t)\mathbf{\hat D} \phi_t+ 
\sigma \mathbf{S}(\mathbf{I_N}\otimes h_1(\theta^{(1)}_t))
(\mathbf{I_N}\otimes \mathsf{P(t)})\dot w_t + 
\dots\\
\lbl{7.12}
&=&g \cl(\theta^{(1)}_t)\mathbf{\hat D} \phi_t+
\sigma \mathbf{S}(\mathbf{I_N}\otimes h_1(\theta^{(1)}_t)\mathsf{P(t)})\dot w_t +\dots. 
\end{eqnarray}
In (\ref{7.12}), $O(|\phi|^2)$ terms are treated as higher
order, because they do not affect exponential stability
of the synchronous solution.
Note how separable coupling translates to the
structure of the phase equation. Matrix $\mathbf{\hat D}$, which carries the
information about the network topology effectively determines the
stability of the synchronous solution. Semipositive matrix
$\mathsf{L}$ enters the factor $\cl(\theta^{(1)})$ (cf. (\ref{ltheta})).
The stability of
(\ref{7.12}) is determined from the homogeneous deterministic system:
\be\lbl{7.13}
\dot\phi_t =g \cl(\theta^{(1)}_t)\mathbf{\hat D} \phi_t,
\ee 
where by $\theta^{(1)}_t$ we mean the first component of the solution 
of deterministic equation (\ref{1.2}$)_0$.
We continue our analysis assuming that $\mathsf{L}>0$, i.e., the coupling
is full. In Section~\ref{generalizations}, we comment on how our results apply to 
partially coupled systems. Thus, $\cl(\mathsf{\theta_t^{(1)}})\ge\alpha>0$
and after changing the independent variable 
 we have
\be\lbl{7.14}
\dot\phi_\tau =g \mathbf{\hat D} \phi_\tau.
\ee 
For exponential stability of synchronous solution,
the symmetric part of $\mathbf{\hat D}$ must be negative definite. 
This motivates
the following definition.
\begin{df}\lbl{df.1}
Matrices from
\be\lbl{diss}
\mathcal{D}=\left\{\mathbf{M}\in\mathcal{K}:\; \mathbf{x^T \hat Mx} <0\; 
\forall \mathbf{x}\in\Re^{N-1}/\{0\}
\right\}
\ee
are called dissipative.
\end{df} \noindent
Thus, we arrive at the first conclusion of this 
Letter: synchronous solution of (\ref{1.2}$)_0$-(\ref{kerD}) is exponentially
stable iff $\mathbf{D}$ is dissipative.
When studying (\ref{1.2}$)_0$-(\ref{kerD}) it is tempting to relate the 
stability of synchronous solution to the spectrum of $\mathbf{D}$.
It is important to realize that it is the spectrum of
$\mathbf{\hat D}$ that is responsible for synchronization.
Remarkably, dissipative matrices admit an explicit characterization.
\begin{thm}\lbl{thm.ql}
$\mathbf{D}\in\mathcal{D}$ iff 
\be\lbl{QLambda}
\mathbf{D=Q\Lambda_0,\; \Lambda_0=S^TS}
\ee
for some $\mathbf{Q}\in\Re^{N\times N}$ with negative definite symmetric part.
\end{thm}
\noindent \pf 
Suppose (\ref{QLambda}) holds. Let $\mathbf{\hat D}=\mathbf{SQS^T}$.
Then
$$
\mathbf{SD}=\mathbf{SQS^TS}=\mathbf{\hat DS}.
$$
Furthermore, for any $\mathbf{x}\in\Re^{N-1}/\{0\}$
\be\lbl{pf1}
(\mathbf{\hat Dx},\mathbf{x})=
(\mathbf{SQS^Tx},\mathbf{x})=(\mathbf{Qy},\mathbf{y}),\;\;
\mathbf{y}=\mathbf{S^Tx},
\ee
where by $(\cdot,\cdot)$ we denote the inner product
in $\Re^{N-1}$. We will use the same notation for the inner product 
in any other Euclidean space used in this Letter, e.g., 
$\Re^N$ and $\Co^N$.
Note that $\mathbf{y}=\mathbf{S^Tx}\ne 0$, because
the rows of $\mathbf{S}$ are linearly independent. Thus, from 
(\ref{pf1}) we conclude
$$
(\mathbf{\hat D x},\mathbf{x})<0 \;\forall \mathbf{x}\in\Re^{N-1}/\{0\},
$$
i.e., $\mathbf{D}\in\mathcal{D}$.

Conversely, suppose $\mathbf{D}\in\mathcal{D}$. Note that on the 
orthogonal complement of $\ker~\mathbf{\Lambda_0}$,
$(\ker~\mathbf{\Lambda_0})^\perp=\one^\perp$,
$\mathbf{\Lambda_0}$ is invertible and define
$\mathbf{Q}\in\Re^{N\times N}$ as follows: for $\mathbf{x}\in\Re^N$
let
\be\lbl{defineQ}
\mathbf{Qx}=\left\{
\begin{array}{cc}
\mathbf{D}(\mathbf{\Lambda_0}\left|_{\one^\perp}\right.)^{-1}\mathbf{x},& 
\mathbf{x}\in\one^\perp,\\
-\mathbf{x},& \mathbf{x}\in\mbox{Span}\{\one\}.
\end{array}
\right.
\ee
We show that the symmetric part of 
$\mathbf{Q}\left|_{\one^\perp}\right.$ (and, therefore,
$\mathbf{Q}$ itself) is negative definite.
For any $\mathbf{z}\in\one^\perp/\{0\}$ there exists 
$\mathbf{x}\in\Re^{N-1}/\{0\}$ such that $\mathbf{z}=\mathbf{\Lambda_0x}$,
because $\mathbf{\Lambda_0}\left|_{\one^\perp}\right.$ is invertible.
Moreover, such $\mathbf{x}$, can be chosen from $\one^\perp/\{0\}$
because $\ker~\mathbf{\Lambda_0}=\mbox{Span}~\{\one\}$. Thus,
\be\lbl{pf3}
(\mathbf{Qz},\mathbf{z})=
(\mathbf{D}(\mathbf{\Lambda_0}\left|_{\one^\perp}\right.)^{-1}\mathbf{\Lambda_0x},
\mathbf{\Lambda_0x})=(\mathbf{Dx},\mathbf{S^TSx})=
(\mathbf{SDx},\mathbf{Sx})=(\mathbf{\hat DSx},\mathbf{Sx})<0.
\ee
Here, we used the fact that $R(\mathbf{\Lambda_0})=\one^\perp$
and $\mathbf{Sx}\neq 0$  
(because $\mathbf{x}\in (\ker~\mathbf{S})^\perp$).  
The combination of (\ref{defineQ}) and (\ref{pf3}) yields
(\ref{QLambda}).\\
$\qed$

Theorem~\ref{thm.ql} gives explicit and 
for separable coupling exhaustive characterization of coupling matrices 
that generate exponentially stable
synchronous solutions. Synchronization is often attributed to systems
with diffusive coupling that are obtained by discretizing elliptic differential
operators or, more generally, differential operators modeling diffusion 
on graphs. In this respect, it is remarkable that Theorem~\ref{thm.ql} 
relates all dissipative  matrices to the discrete Laplacian
\be\lbl{Lambda_0}
\mathbf{\Lambda_0}=\mathbf{S^TS}=\left(\begin{array}{cccccc}
1 &-1 & 0& \dots &0&0 \\
-1 & 2& -1 & \dots& 0&0\\
\dots&\dots&\dots&\dots&\dots&\dots\\
0 &0 & 0& \dots &-1 &1
\end{array}
\right).
\ee
This is consistent with the common interpretation
of the Laplacian as a prototypical elliptic operator. 
The explicit characterization of dissipative matrices by 
(\ref{QLambda}) is the second result of this Letter.
Before turning to the question of the robustness to noise we
state two corollaries of Theorem~\ref{thm.ql}. The first
corollary gives a convenient computational formula for
$\mathbf{\hat D}$, while second one characterizes the
spectrum of a dissipative matrix.
\begin{cor}\lbl{Q1}
For $\mathbf{D}\in\mathcal{D}$, $\hat D=\mathbf{S}Q\mathbf{S^T}$, 
where $\mathbf{Q}$ satisfies (\ref{QLambda}).
\end{cor}
\begin{cor}\lbl{Q2}
If $\mathbf{D}\in\mathcal{D}$ then $\mathbf{D}$ has
a zero eigenvalue of multiplicity $1$. 
All nonzero eigenvalues of $D$ are real and negative.
\end{cor}
\noindent \pf 
Matrix $\mathbf{D}$ has a simple zero eigenvalue by (\ref{kerD}).
Suppose $\lambda\in\Co$ is a nonzero eigenvalue of $D$ and 
$\mathbf{x}\in\Co^n$ be a corresponding eigenvector
\be\lbl{eigenvalue}
\mathbf{Q\Lambda_0x}=\lambda\mathbf{x}.
\ee
 Note that
$x\notin \mbox{Span}\{\one\}$.
By multiplying both sides of (\ref{eigenvalue}) by 
$\mathbf{\Lambda_0x}\ne 0$,
we have
\be\lbl{eig1}
(\mathbf{Q\Lambda_0x},\mathbf{\Lambda_0x})=
\lambda (\mathbf{x},\mathbf{\Lambda_0 x})=\lambda 
(\mathbf{Sx},\mathbf{Sx}).
\ee
The scalar product multiplying $\lambda$ on the right hand side of 
(\ref{eig1}) is positive, while 
$(\mathbf{Q\Lambda_0x},\mathbf{\Lambda_0 x})$ is negative.
Therefore, $\lambda <0$.
$\qed$

Having understood the mechanism for synchronization in the deterministic 
network (\ref{1.2}$)_0$. 
We now turn to the question of robustness of the synchronous regime to noise. 
To this end,
we return to (\ref{7.12}). For the remainder of this Letter, 
$\mathbf{D}\in\mathcal{D}$.
For small $\sigma>0$, 
on a finite time interval solution
of (\ref{7.12}) can be expanded as
\be\lbl{i}
\phi_t=\bar\phi_t+\sigma\tilde\phi_t+\dots,
\ee
where deterministic function $\bar\phi_t$ solves (\ref{7.13}) 
(cf. Theorem~2.2, \cite{FW}). 
Since $\phi\equiv 0$ is
an exponentially stable solution of (\ref{7.13}), we take $\bar\phi_t\equiv 0$.
The leading order correction $\sigma\tilde\phi_t$ is a Gaussian
process, which for small $\sigma>0$ approximates $\phi_t$ 
on a finite interval of time. This specifies the scope of applicability
of our analysis. In particular, we are not concerned with large deviation
type effects which become relevant on much longer timescales.

From (\ref{7.9}) we have
\be\lbl{ii}
\theta_t^{(1)}=t+O(\sigma).
\ee
By plugging (\ref{i}) and (\ref{ii}) in (\ref{7.12}), we have
\be\lbl{iii}
\dot{\tilde{\phi_t}} =g \cl(t)\mathbf{\hat D} \tilde\phi_t+ 
\mathbf{S}(\mathbf{I_N}\otimes h_1(t)\mathsf{P(t)})\dot w_t 
+\dots.
\ee
After changing time to $\tau=\int_0^t\cl(s)ds$ and ignoring higher order terms, 
(\ref{iii}) is rewritten as
\be\lbl{iii}
\dot{\tilde{\phi_\tau}} =g\mathbf{\hat D} \tilde\phi_\tau+ 
\mathbf{S}(\mathbf{I_N}\otimes h_1(\tau)\mathsf{\tilde P(\tau)})\dot w_\tau,\; 
h(\tau):={h_1(t(\tau))\over\sqrt{\cl(t(\tau))}},\;\;
\mathsf{\tilde P(\tau)}:=\mathsf{P(t(\tau))}.
\ee
The solution of (\ref{iii}) subject to initial condition $\tilde\phi_0=0$ is 
a Gaussian random process with zero mean and covariance matrix 
(cf. \S 5.6, \cite{KS})
\begin{eqnarray}\nonumber
\cov\tilde\phi_\tau&=&\int_0^\tau e^{g(\tau-s)\mathbf{\hat D}}
\mathbf{S}(\mathbf{I_N}\otimes h(s)\mathsf{\tilde P(s)})
(\mathbf{I_N}\otimes h(s)\mathsf{\tilde P(s)})^T\mathbf{S^T}
e^{g(\tau-s)\mathbf{\hat D^T}} ds\\
\lbl{iv}
&=& \int_0^\tau \tilde h^2(s) e^{g(\tau-s)\mathbf{\hat D}}\mathbf{\Lambda}
e^{g(\tau-s)\mathbf{\hat D^T}} ds,
\end{eqnarray}
where $\mathbf{\Lambda}:=\mathbf{SS^T}$
and $\tilde h^2(t)=\mathbf{I_N}\otimes 
(h(t)\mathsf{\tilde P(t)\tilde P(t)^T} h^T(t))$ is a nonnegative scalar
function. Using standard properties of trace and
the mean value theorem, from (\ref{iv}) we have
\begin{eqnarray}\nonumber
\tr\cov\tilde\phi_\tau&=&\int_0^\tau \tilde h^2(u)\tr\{\mathbf{\Lambda} 
e^{g(\tau-u)\mathbf{\hat D^s}}\}du\\
\nonumber
&=&\tilde h^2(\zeta(\tau))\tr\{ \mathbf{\Lambda} 
e^{g\tau\mathbf{\hat D^s}} \int_0^\tau e^{-g u\mathbf{\hat D^s}} du\} \\
\nonumber
&=&\tilde h^2(\zeta(\tau)) \tr\left\{-\mathbf{\Lambda} 
g^{-1}(\mathbf{\hat D^s})^{-1}
\left[\mathbf{I_{N-1}}-e^{g\tau\mathbf{\hat D^s}}\right]\right\}\\
\lbl{v}
&=&\mu(\tau){\kappa(\mathbf{D})\over g} + O(e^{-c_1\tau}),
\end{eqnarray}
where continuous function $\zeta(\tau)$ is due to the 
application of the mean value theorem, 
$\mu(\tau):=\tilde h^2(\zeta(\tau))$, 
$\mathbf{\hat D^s}:=\mathbf{\hat D +\hat D^T}$, 
and 
\be\lbl{viii}
\kappa (\mathbf{D})=\tr\{-\mathbf{\Lambda} (\mathbf{\hat D^s})^{-1}\}.
\ee
Note that $0\le \mu(\tau)\le M$ is uniformly bounded.
Define the average variance of the variables 
$\mathsf{\phi^{(k)}_t},\; k=1,2,\dots,N$ as
\be\lbl{ix}
\ovar\phi_t={1\over N-1}\sum_{k=1}^{N-1} \var\mathsf{\phi^{(k)}_t}=
{1\over N-1}\tr\cov \phi_t.
\ee
Equation (\ref{v}) yields an important estimate for the network variability
\be\lbl{x}
\ovar\phi_\tau\approx\sigma^2\ovar\tilde\phi_\tau
=\sigma^2\mu(\tau){\kappa(\mathbf{D})\over g (N-1)} + O(e^{-c_1\tau}),
\ee
where $c_1$ is a positive constant.
Nonnegative function $\mu(\tau)$ reflects 
the properties of the local system
such as geometric properties of the limit cycle
and matrix $\mathsf{P}$ multiplying stochastic term, 
whereas $\kappa(\mathbf{D})$ captures network topology.
For a network of fixed size, $\ovar\phi_t$ can be made arbitrarily small by
taking large $g$. Moreover, by Chebyshev's inequality, for any $\delta>0$,
$$
\P\left\{\left|\mathsf{\theta^{(j)}_t-\theta^{(i)}_t}\right|>\delta\right\}\le 
{\sigma^2 M N \kappa(\mathbf{D})\over \delta^2 g}\to 0\;\mbox{as}\;g\to\infty,
$$
i.e., for strong coupling the phases of individual oscillators
can be localized within arbitrarily narrow
bounds. The control of the coherence by varying the coupling strength is more
effective in networks with smaller $\kappa(\mathbf{D})$. Thus, (\ref{x}) shows
explicitly the factors controlling the coherence in the presence of noise. 
Moreover, $\kappa(\mathbf{D})$ quantifies the contribution of the
network topology to the stability of the synchronous state.
This is the third conclusion of this Letter.

What features of the network topology are captured
by $\kappa(\mathbf{D})$? We first go over the ingredients of the formula
for $\kappa(\mathbf{D})$ (cf. (\ref{viii})).
Matrix $\mathbf{\Lambda}$ is a Laplacian:
\be\lbl{Lambda}
\mathbf{\Lambda}=\mathbf{SS^T}=\left(\begin{array}{cccccc}
2 &-1 & 0& \dots &0&0 \\
-1 & 2& -1 & \dots& 0&0\\
\dots&\dots&\dots&\dots&\dots&\dots\\
0 & 0 & 0& \dots &-1 &2
\end{array}
\right).
\ee
Unlike $\mathbf{\Lambda_0}$, $\mathbf{\Lambda}$ is nonsingular. The following
examples show that $\kappa(\mathbf{D})$ can change by many times for networks with
different topologies. Let  $\mathbf{D_1}$ be as in (\ref{all}) and 
$\mathbf{D_2}=-\mathbf{\Lambda_0}$ (cf. (\ref{Lambda_0})).
$\mathbf{D_1}$ and $\mathbf{D_2}$ are coupling matrices 
corresponding to the graphs 
modeling all-to-all and nearest-neighbor interactions in the network. 
By direct verification,
\be\lbl{hat}
\mathbf{\hat D^s_1}=-2N~\mathbf{I_{N-1}}\;
\mbox{and}\;\mathbf{\hat D^s_2}=-2\mathbf{\Lambda}.
\ee
By plugging the explicit expressions 
(\ref{hat}) in (\ref{viii}), 
we find 
\be\lbl{kappa}
\kappa(\mathbf{D_1})=1+O(N^{-1})
\;\mbox{and}\;
\kappa(\mathbf{D_2})={N-1\over 2}.
\ee
Note that the all-to-all topology features a significant reduction in $\kappa$
compared to the nearest-neighbor coupling. This reduction is proportional to the
ratio of the degrees of the corresponding graphs: 
$2$ - for the nearest-neighbor and $(N-1)$ - for the all-to-all coupling.
Thus, (\ref{x}) suggests that networks with the higher density 
of connections are more
robust to noise. 

Equation (\ref{x}) estimates the variability of the phase differences, revealing
the main factors contributing to robustness of synchrony to noise.
If $\mathbf{D}$ is symmetric, Equation (\ref{x}) can also be used
to estimate the variability of the phase variables $\mathsf{\theta_t^{(i)}}$
themselves. We follow the method used in \cite{M09} for a related problem.
First, we derive the equation for the average phase of the coupled system
$$
\mathsf{\bar\theta}=\eta^T\theta,\;\eta=N^{-1}\one.
$$  
By multiplying both sides of (\ref{7.9}) by $\eta^T$, we have
\be\lbl{averagephase}
\mathsf{
\dot{\bar{\theta_t}}=1+\sigma h_1(t)P(t)\dot X_t+\dots, \; 
\dot X=}{1\over N}\sum_{i=1}^N \mathsf{\dot w^{(i)}_t.
}
\ee
Here, we used the following approximations
$$
V_N(\theta)=\cl(\mathsf{\theta^{(1)}_t})\mathbf{I_N}+O(|\phi|)
\quad\mbox{and}\quad \mathsf{\theta^{(1)}_t}=t+O(\sigma).
$$
As a linear combination of independent Gaussian processes 
$\mathsf{w_t^{(i)}}$,
$\mathsf{X(t)}$ is distributed as 
$N^{-1/2} \mathsf{w_t}$, where $\mathsf{w_t}$ is a $n-$dimensional
Brownian motion.
Thus,
\be\lbl{averagephase1}
\mathsf{
\dot{\bar{\theta_t}}=1+{\sigma h_1(t)P(t)\over\sqrt{\mathrm{N}}}\dot w_t+\dots
}
\ee
and 
\be\lbl{vartheta}
\var\mathsf{\bar\theta_t}\approx
{\sigma^2\over N} \int_0^t \mathsf{h_1(s)P(s)P(s)^Th_1(s)^T}ds
\le {\sigma^2C_1 T\over N},\; t\in [0,T],
\ee
where $C_1>0$ does not depend on $N$ and $T$.
Next, by noting as in \cite{M09} that each phase variable 
$\mathsf{\theta^{(i)}}$ can be represented as a linear combination
of the average phase $\mathsf{\bar\theta}$ and phase differences
$\phi^{(1)},\phi^{(2)},\dots,\phi^{(N)}$, we estimate
$\var\mathsf{\theta^{(i)}}$ using (\ref{vartheta}) and (\ref{x}).
Omitting further details, which are the same as in step 
{\bf 3.} of the proof of Theorem 3.1 in \cite{M09}, we state  
the final result
\be\lbl{varthetai}
\max_i\sup_{t\in[0,T]}\var\mathsf{\theta^{(i)}_t}\le \sigma^2\left({C_2T\over N}
+{C_3N\kappa(\mathbf{D})\over g}\right), 
\ee
where $C_2$ and $C_3$ are positive constants independent 
from $N, g$ and $T$. 
The first term on the right hand side of (\ref{vartheta})
can be made arbitrarily small by increasing $N$, while the
second term decreases for increasing $g$. Therefore,
in large networks, the effects of noise on oscillations
can be controlled by varying the strength of coupling.
The two terms on the right hand side of (\ref{vartheta})
represent two main ingredients of the mechanism of denoising:
the first term is due to the averaging of statistically independent
stochastic forces acting on connected local systems, whereas
the second term reflects the dissipativity of the coupling.
The latter is a critical property of the coupling operator
that underlies both synchronization and denoising in coupled
networks. 

The analysis of the phase equations above produced a necessary and sufficient 
condition for synchronization in systems with separable coupling and gave a
compact explicit estimate for the spread of phases of coupled stochastic oscillators.
For the phase equations to be valid, the trajectory of the coupled system must
remain close to the limit cycle. To complete the analysis,
we consider the system of equations for $\rho_t=(\mathsf{\rho_t^{(1)}, \rho_t^{(2)},\dots,
\rho_t^{(N)}})$. The derivation of the system for $\rho$ is completely analogous 
to that for $\theta_t$ (cf. (\ref{7.9})). We omit the details and state
the final result
\be\lbl{rho}
\dot \rho_t=\left[A(t)+g D(t)\right]\rho_t
+\sigma h_2(t)P(t) \dot w_t+\dots,
\ee
where $A(t)=\mathbf{I_N}\otimes\mathsf{A(t)}$, 
$D(t)=\mathbf{D}\otimes(\mathsf{Z^T(t)LZ(t)})$. Matrices $\mathsf{Z(t)}$
and $\mathsf{A(t)}$ are defined in (\ref{7.2}) and (\ref{7.5}) respectively.
By Vazhevski's inequality \cite{Demidovich}, the combination of (\ref{7.5a}),
$\mathbf{D}\in\mathcal{D}$, and $\mathsf{L}\ge 0$ implies  exponential stability 
of the equilibrium at $\rho=0$ in (\ref{rho}$)_0$. Therefore, on finite time
intervals with overwhelming probability, $\left|\rho_t\right|$ remains small,
provided $\left|\rho_0\right|$ and $\sigma>0$ are sufficiently small.
This justifies the phase reduction that we used above. 
Note that this conclusion holds for both full and partial coupling. 

\section{Generalizations}
\lbl{generalizations}
\setcounter{equation}{0}
The analysis of this Letter admits several important generalizations.

A) Partial coupling. If the coupling is partial, nonnegative function
$\cl(\cdot)$ in (\ref{iii}) in general takes zero values. Dealing
with the degeneracies in (\ref{iii}) requires additional care.
For a common in applications case when $\cl(\cdot)$ has isolated
zeros, with technical modifications one can get a qualitatively
similar estimate to (\ref{x}).

B) Time-dependent coupling.  Our analysis remains unchanged if
instead constant $\mathsf{L}$ one uses a bounded measurable function of
time. The definition of the full coupling is then modified to
$\mathsf{x^TL(t)x}\ge\alpha\mathsf{x^Tx}$ for some $\alpha >0$ and
$\forall\mathsf{x}\in\Re^n/\{0\}$ uniformly in $t\ge 0$.
Likewise, $\mathbf{D}$ can be taken time-dependent as long as 
$\mathbf{D}(t)\in\mathcal{D}$ for all $t$. 
In this case, exponential stability of $\phi\equiv 0$ follows
from (\ref{7.14}) if we require that all eigenvalues of
$\mathbf{\hat D^s}(t)=\mathbf{\hat D}(t)+\mathbf{\hat D^T}(t)$
are negative and bounded away from zero uniformly in $t$:
$$
\mathbf{x^T\hat D^s(t)x}=
\mathbf{x^T\hat D(t) x}\le -\gamma \mathbf{x^T x}\;\;
\forall \mathbf{x}\in\Re^{N-1}/\{0\},\; t\ge 0
$$
for some $\gamma>0$. 
By Theorem~\ref{thm.ql}, such matrices can be written as
$
\mathbf{D}=\mathbf{Q(t)\Lambda_0},
$
where $\mathbf{Q(t)}$ is such that
$$
\mathbf{x^T Q(t)x}\le -\tilde\gamma \mathbf{x^Tx}\;\;
\forall \mathbf{x}\in\Re^{N}/\{0\},\; t\ge 0
$$
for some $\tilde\gamma>0$.
Also, in the time-dependent case, one can get a 
slightly weaker but qualitatively similar estimate on $\ovar\phi_t$ 
(cf. (\ref{x})).

C) Nonlinear coupling. The analysis can be  extended to 
the systems with nonlinear coupling of the following form
\be\lbl{non.1}
\mathsf{
\dot x^{(i)}=f(x^{(i)})}+g\sum_{j=1}^N
\mathsf{\mathbf{d_{ij}(t)} \mathsf{\tilde L}(x^{(i)}),
}\;
i=1,2,\dots, N,
\ee
where $\mathbf{D(t)}=(\mathbf{d_{ij}(t)})\in\mathcal{K}$ 
for every $t\ge 0$ and
$\mathsf{\tilde L}:\Re^n\to\Re^n$ is a smooth function.
Since $\mathbf{D(t)}\in\mathcal{K}$,
(\ref{non.1}) can be rewritten as
\be\lbl{non.3}
\mathsf{
\dot x^{(i)}= f(x^{(i)})}+g\sum_{j=1}^N\mathsf{\mathbf{d_{ij}(t)}
\left(\tilde L(x^{(j)})-\tilde L(x^{(i)})\right), }
\; i=1,2,\dots, N.
\ee
Suppose $\mathsf{x=\xi(t)}$ is a periodic solution of the local
system $\mathsf{\dot x= f(x)}.$ Then $x_p=\one\otimes\mathsf{\xi(t)}$
solves (\ref{non.3}). In a neighborhood of $x_p$, (\ref{non.3})
can be rewritten using Taylor's formula
$$
\mathsf{
\dot x^{(i)}= f(x^{(i)})}+g\sum_{j=1}^N\mathsf{\mathbf{ d_{ij}(t)} \mathsf{L(t)}
(x^{(j)} -x^{(i)})+ O(\max_{k,l}| x^{(k)}-x^{(l)}|^2), \; 
}
\;\; i=1,2,\dots, N,
$$
or, equivalently,
\be\lbl{non.4}
\dot x= f(x)+g(\mathbf{D(t)}\otimes\mathsf{L(t)})x+\dots,
\;\mbox{where}\;
\mathsf{L(t)}:={\partial \mathsf{\tilde L(\xi(t))}\over\partial \mathsf{x}}.
\ee
Equation (\ref{non.4}) is now in the form, for which the analysis
of Section~2 applies (cf. B) above).

D) Nonperiodic attractors. 
Moving coordinate systems similar to the one used in this Letter
can be introduced in the vicinity of locally invariant sets of more 
general nature. For example, motions along certain attracting normally 
hyperbolic slow manifolds admit a similar description 
(cf. \cite{hale_odes, wiggins}). Thus, our analysis can 
be adopted to study synchronization in a more general setting. 
Furthermore, in the case when the attractor of the local system
is periodic and synchronization takes place, the analysis of Section~2
yields a precise description 
of the asymptotic behavior of trajectories of the coupled system. 
Specifically, the attractor  of the coupled system includes a periodic
orbit that is a direct product of the periodic orbits
of the local systems. If one is only concerned with synchronization
and does not aim at describing the asymptotic behavior of the coupled system,
then the transformation to moving coordinates is not needed. To outline
the analysis for this case, let $\mathsf{x=\xi(t)}$ be a solution 
of the local system. This solution is not necessarily periodic.
Instead, we assume  that
$\mathsf{x=\xi(t)}$ does not leave a bounded domain in $R^{n}$. 
Then $\xi(t)=\mathbf{1_N}\otimes\mathsf{\xi(t)}$ solves the 
coupled system  
\be\lbl{C.1}
\dot x=f(x)+g(\mathbf{D}\otimes\mathsf{L})x, 
x=(\mathsf{x^{(1)},x^{(2)},\dots, x^{(N)}})\in\Re^{Nn}.
\ee
Linearization about $x=\xi(t)$ yields
\be\lbl{C.2}
\dot x=(\mathbf{I_N}\otimes\mathsf{A(t)}+g(\mathbf{D}\otimes\mathsf{L}))x
+\dots, \quad \mathsf{A(t)}={\mathsf{\partial f(\xi(t))}\over\mathsf{ \partial x}}.
\ee
By multiplying both sides of (\ref{C.2}) by $S=\mathbf{S}\otimes\mathsf{I}$,
we get the equation of $y=Sx$:
\be\lbl{C.3}
\dot y=(\mathbf{I_{N-1}}\otimes\mathsf{A(t)}+
g(\mathbf{\hat D}\otimes\mathsf{L}))y
+\dots.
\ee  
Asymptotic stability of $y\equiv 0$ implies synchronization
for (\ref{C.1}). A sufficient condition for asymptotic stability of the
trivial solution of (\ref{C.3}) is that the eigenvalues
of symmetric matrix
\be\lbl{C.4}
B=\mathbf{I_{N-1}}\otimes\mathsf{A^s(t)}+g\mathbf{\hat D^s}\otimes\mathsf{L}
\ee
are negative and bounded from zero uniformly in $t\ge 0$. Since
$|\mathsf{A^s(t)}|$ is bounded, the desired property for $B$
for large $g$ follows from $\mathbf{D}\in\mathcal{D}$ and
$\mathsf{L}$ being symmetric positive definite. Thus, we get 
a sufficient condition for synchronization for the full coupling.
In the partial coupling case, $\mathbf{\hat D^s}\otimes L$
has $N\times (n-\mbox{rank}~(\mathsf{L}))$ zero eigenvalues and one 
has to make sure that they do not give rise to negative 
eigenvalues of $B$. The condition for this can be obtained
from the well-known formulas for the perturbations of the eigenvalues
of symmetric matrices (cf. Appendix in \cite{gelfand}).
A more complete analysis of synchronization for the partial
coupling case will be given elsewhere \cite{inprep}.

\section{Numerical example} \lbl{numerical}
\setcounter{equation}{0}

\begin{figure}
\begin{center}
{\bf a}\epsfig{figure=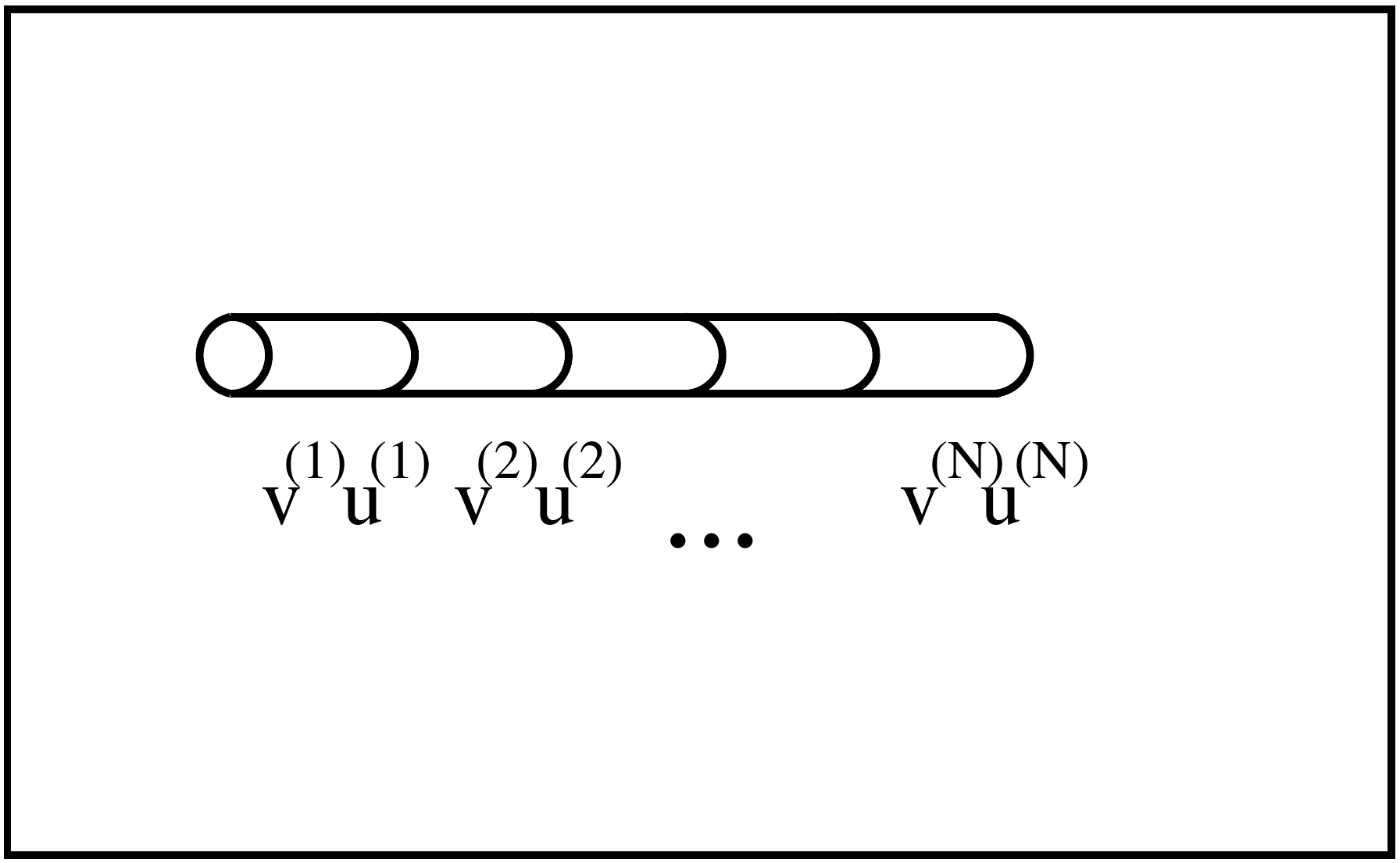, height=2.0in, width=2.5in, angle=0}
\qquad
{\bf b}\epsfig{figure=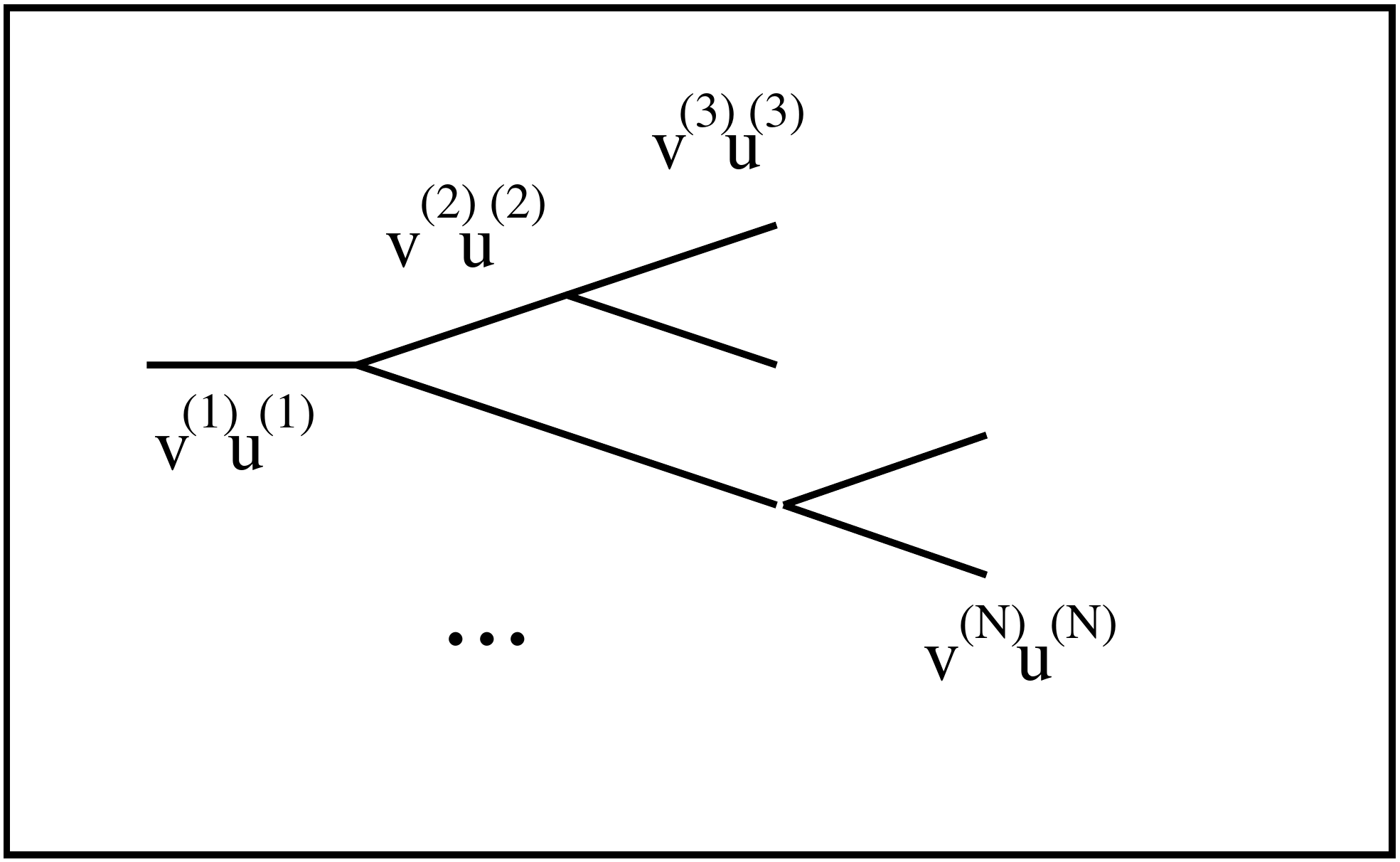, height=2.0in, width=2.5in, angle=0}
\end{center}
\caption{Schematic representation of spatial structure of a compartmental 
model:
(a) linear cable, (b) branched cable. Dynamic variables $v^{(i)}$ and 
$u^{(i)}$, $i=1,2,\dots,N$, approximate voltage and calcium concentration
in each compartment.
}\label{f.1}
\end{figure}

To illustrate the analytical results  of this Letter,
we use a nondimensional model of a pacemaker neuron from
\cite{MC04}:
\begin{eqnarray}\lbl{n.1}
\epsilon \dot v_t^{(i)}  &=&  g_1(v_t^{(i)})\left(E_1-v_t^{(i)}\right)+
g_2(u_t^{(i)})\left(E_2-v_t^{(i)}\right)
+\bar g_3\left(E_3-v^{(i)}\right) 
+I_v^{(i)}+\sigma \dot w_t^{(i)},\\ 
\dot u_t^{(i)}&=&\omega\left( g_1(v_t^{(i)})
\left(E_1-v_t^{(i)}\right)-{u_t^{(i)}\over \tau}\right)
+I_u^{(i)}, \quad i=1,2,\dots,N.
\lbl{n.2}
\end{eqnarray}
Dynamic variables $v^{(i)}$ and $u^{(i)}$ represent
membrane potential and calcium concentration in a given compartment
of an axon or a dendrite of a neural cell. The compartments are sufficiently
small so that the membrane potential and calcium concentration can be
assumed constant throughout one compartment (Fig.~\ref{f.1}a). 
Terms on the right hand side
of the voltage equation (\ref{n.1}) model ionic currents:
a calcium current, a calcium dependent potassium current, and 
a small leak current. In addition, small white noise is added
to account for random synaptic input or other fluctuations.
The equation for calcium concentration (\ref{n.2}) takes into account
calcium current and calcium efflux due to calcium pump.
The ionic conductances are sigmoid functions of the voltage 
and calcium concentration
\begin{eqnarray}\lbl{n.3}
g_1(v)&=& {\bar g_1\over 2} \left(1+\tanh \left({v-a_1\over
a_2}\right)\right),\\
\lbl{n.4}
g_2(u) &=& {\bar g_2 u^4 \over u^4 +a_3^4}.
\end{eqnarray} 
We briefly comment on the meaning of the model parameters:
$\bar g_{1,2,3}$ and $E_{1,2,3}$ stand for maximal conductances
and reversal potentials of the corresponding ionic currents;
$a_{1,2,3}$ are constants used in the descriptions of activation
of calcium and calcium dependent currents; $\omega$ and $\epsilon$
are certain constants that come up in the process of nondimesionalization
of the conductance based model of a dopamine neuron
(see \cite{MC04} for details).
The values of the parameters that were used in our simulations 
are given in the appendix to this Letter.

The coupling terms 
\begin{eqnarray}\lbl{n.5}
I_v^{(i)} &=&
g\sum_{j=0}^N {\mathbf d_{ij}} (v_t^{(j)}-v_t^{(i)}),\\
\lbl{n.6}
I_u^{(i)} &=&
\delta\sum_{j=0}^N {\mathbf d_{ij}} (u_t^{(j)}-v_t^{(i)})
\end{eqnarray}
model electrical current and calcium diffusion between adjacent
compartments respectively. In case of a linear cable geometry
of an axon (dendrite) shown in Fig.~\ref{f.1}a, $\mathbf{D}=
(\mathbf{d_{ij}})$ is the matrix corresponding to the 
nearest-neighbor coupling (cf. (\ref{Lambda_0})). For branched
dendrites (see Fig.~\ref{f.1}b), $\mathbf{D}$ may have a more 
complex structure. In either case, the structure of $\mathbf{D}$ reflects 
the geometry of the neuron. By combining this information, we
obtain a model in the form of (\ref{1.2}), (\ref{sep}),
with
\be\lbl{n.7}
\mathsf{L}=
\left(
\begin{array}{cc} 1 & 0\\
                  0 & \delta_1 \end{array}
\right)\quad \mbox{and}\quad
\mathsf{P}=
\left(
\begin{array}{cc} 1 & 0\\
                  0 & 0 \end{array}
\right),
\ee
where $\delta_1=g^{-1}\delta$.  The coupling is full rank. If we 
disregard calcium diffusion, i.e., set $\delta=0$, the coupling
becomes partial with $\mathsf{L}=\mbox{diag}~(1,0)$.
System of equations (\ref{n.1}) and (\ref{n.2}) with $\delta=0$
admits an alternative interpretation. One can view $v^{(i)}$
and $u^{(i)}$ as the membrane potential and calcium concentration
of Cell $i$ in a neuronal population. The coupling is
due to the current through the gap-junctions between adjacent cells.
In this case, $\mathbf{D}$ reflects network connectivity.
If gap-junctional conductance depends on voltage or calcium 
concentration, the coupling is nonlinear as in (\ref{non.1}).
If the gap-junctions permit ions in one direction, 
coupling matrix $\mathbf{D}$ is not symmetric. 

\begin{figure}
\begin{center}
{\bf a}\epsfig{figure=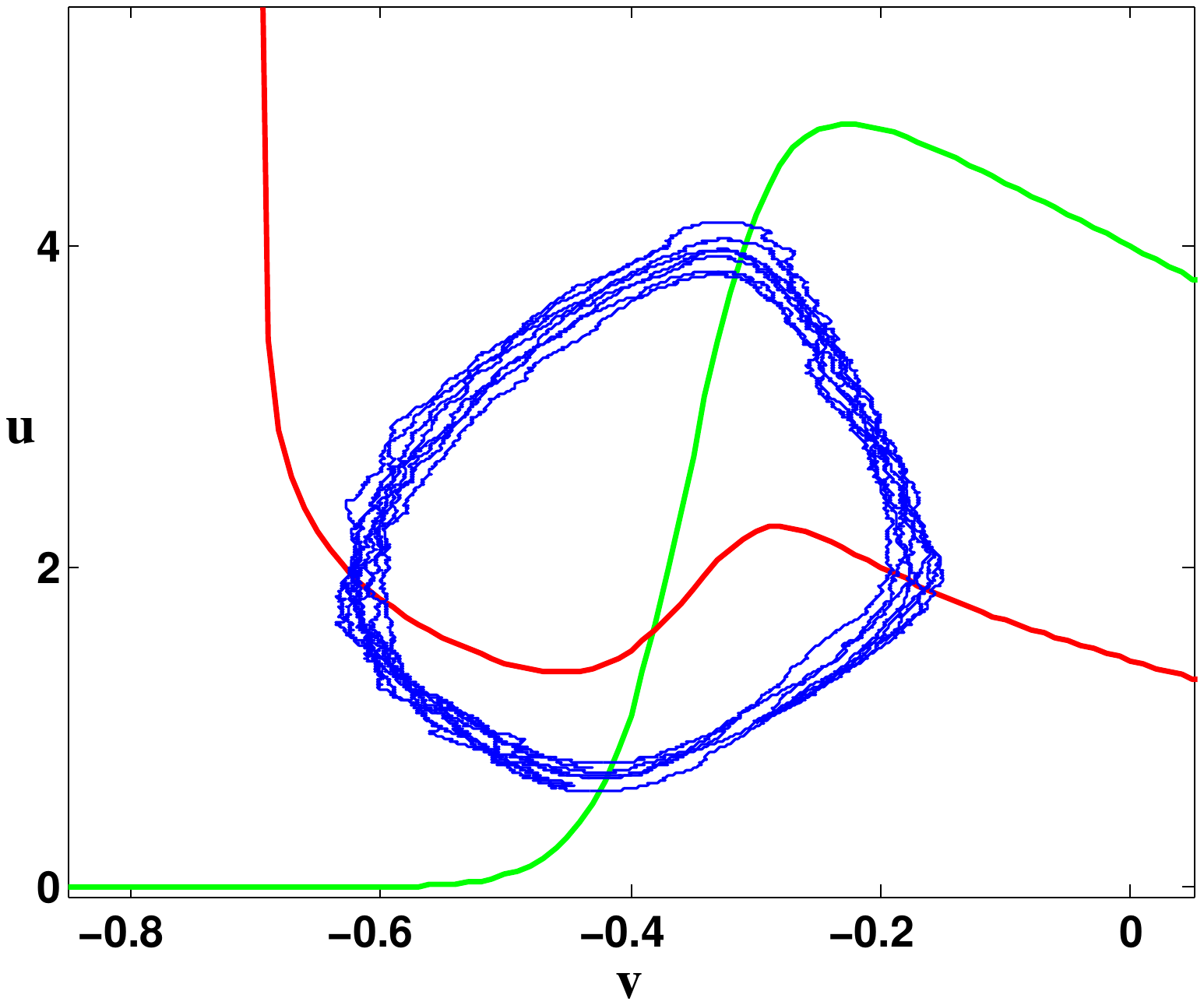, height=2.0in, width=2.5in, angle=0}
{\bf b}\epsfig{figure=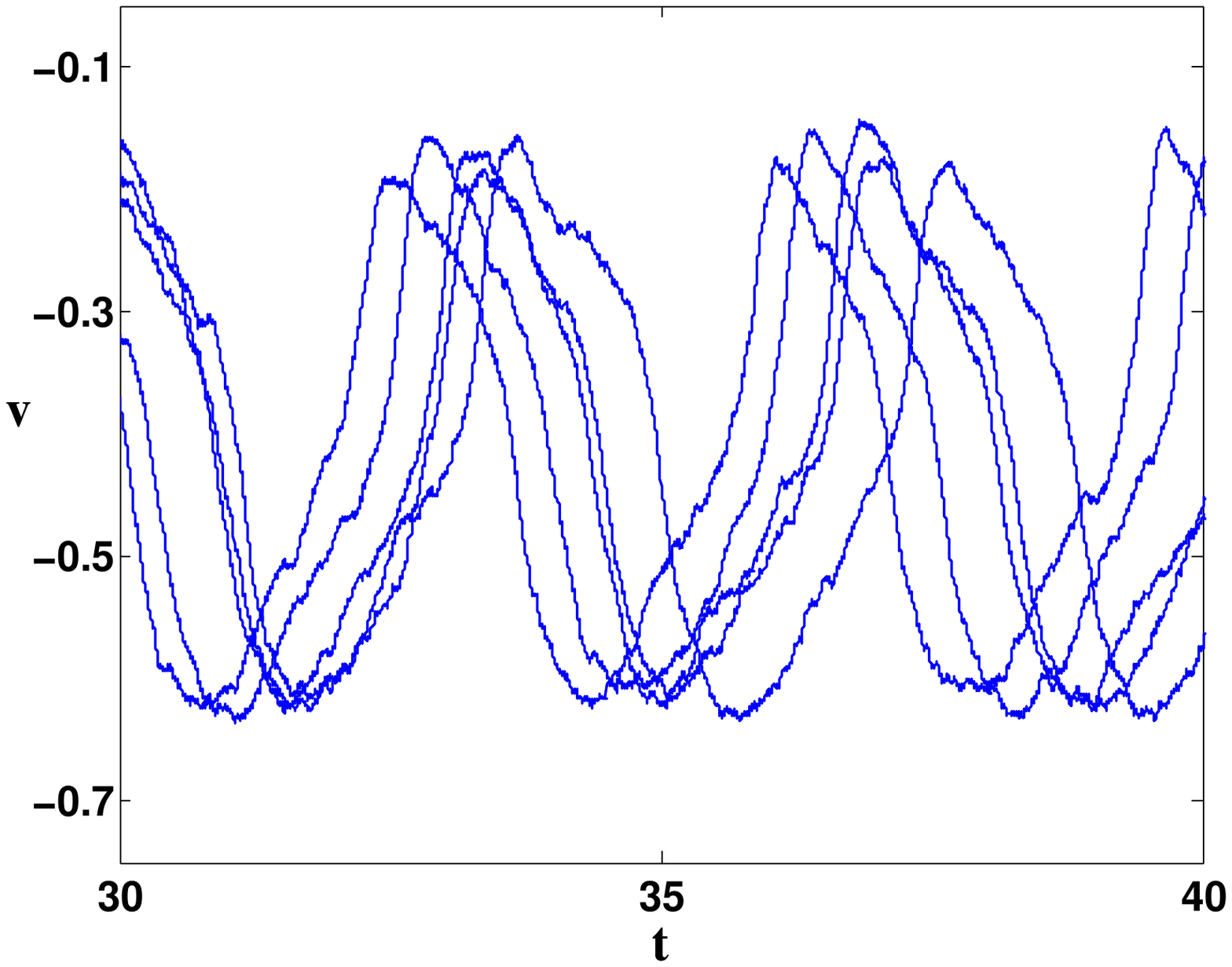, height=2.0in, width=2.5in, angle=0}
{\bf c}\epsfig{figure=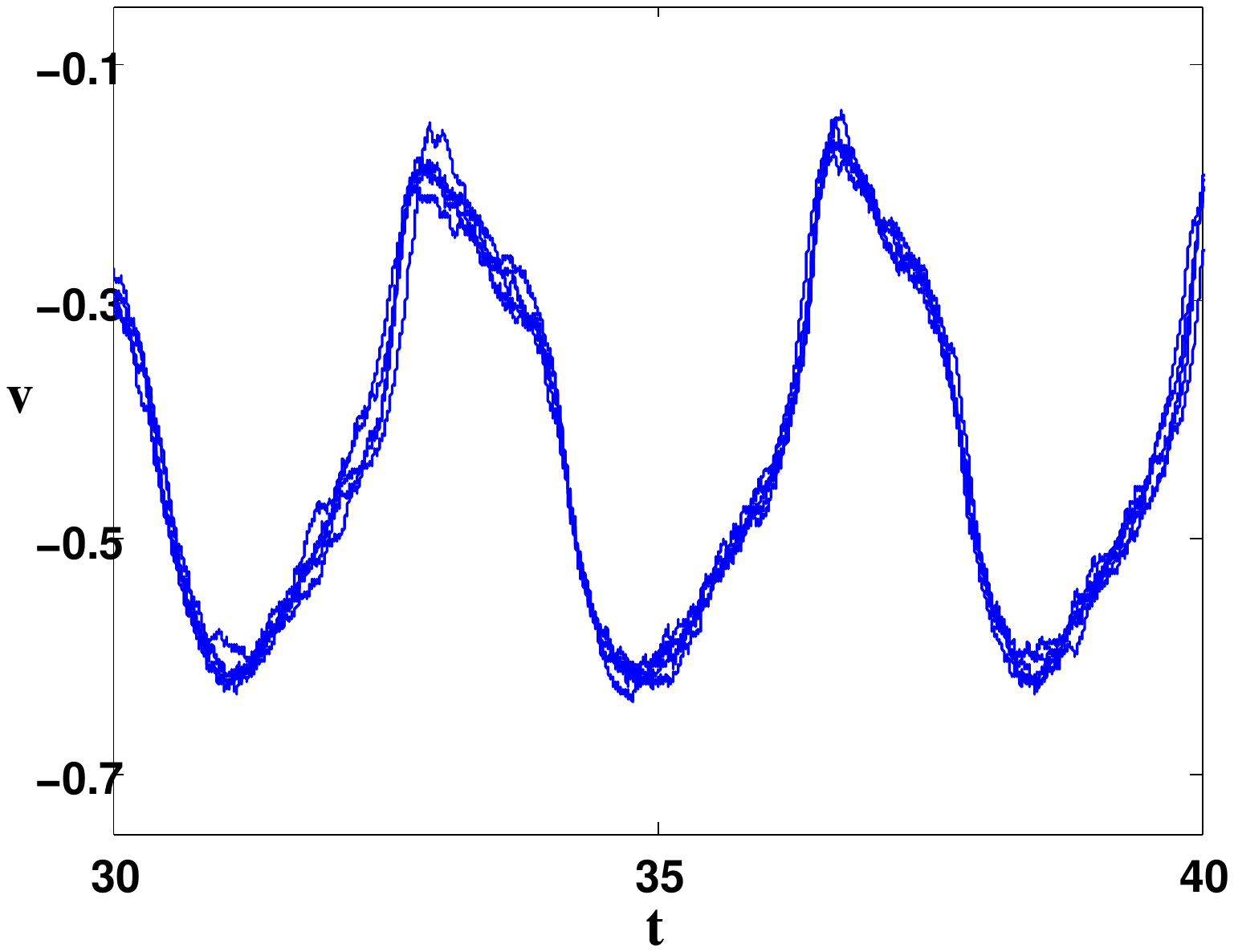, height=2.0in, width=2.5in, angle=0}
{\bf d}\epsfig{figure=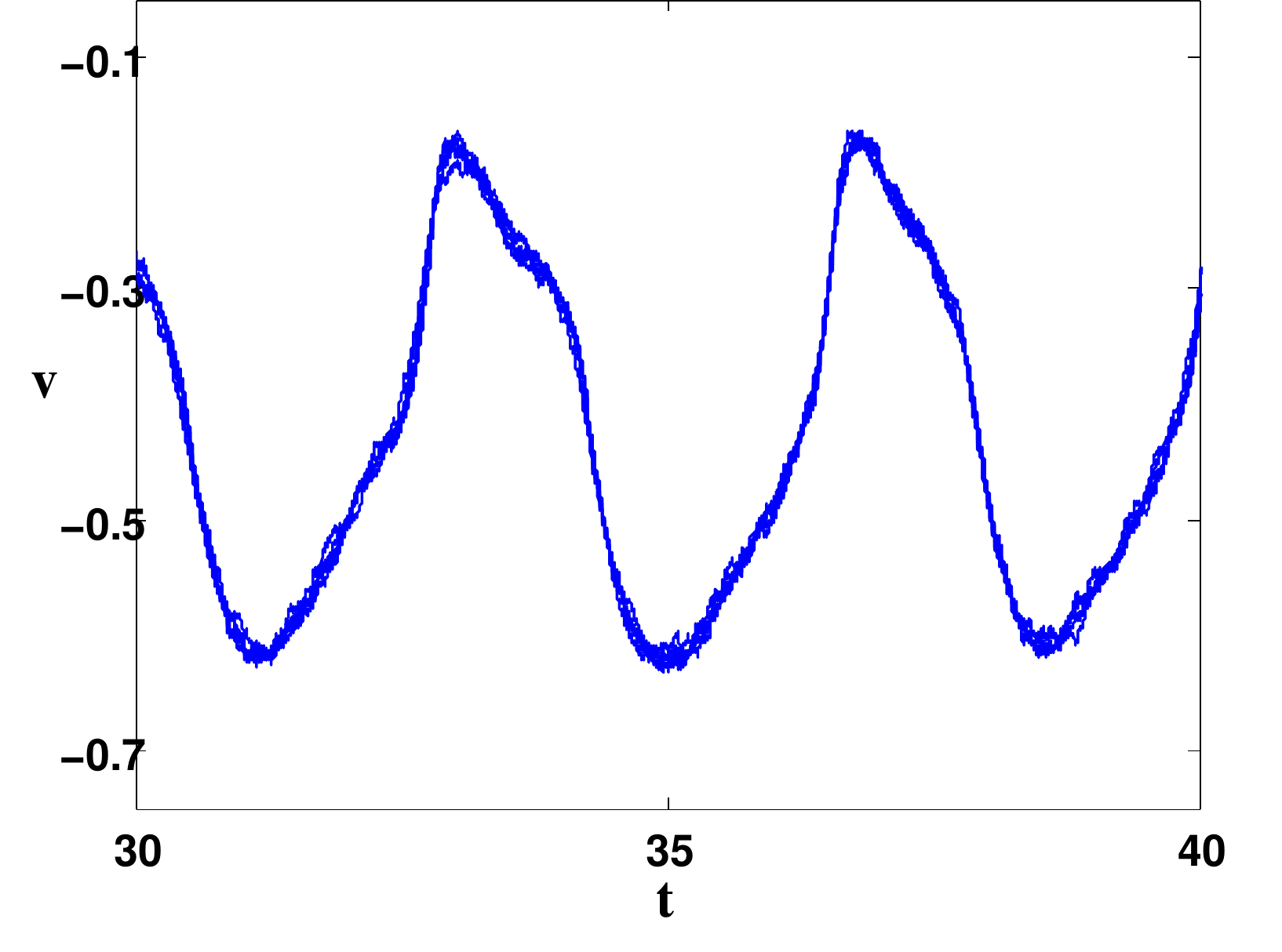, height=2.0in, width=2.5in, angle=0}
{\bf e}\epsfig{figure=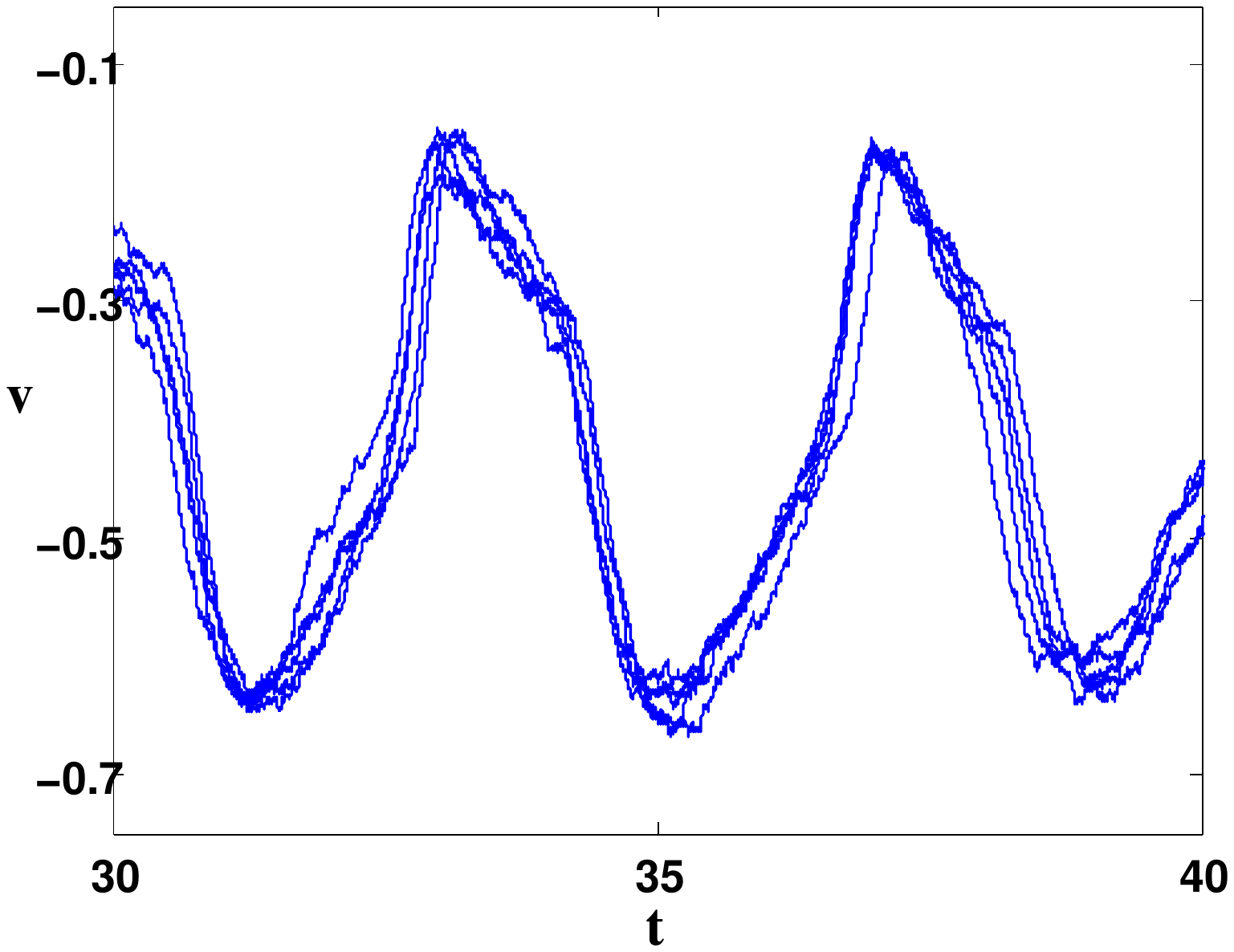, height=2.0in, width=2.5in, angle=0}
\end{center}
\caption{Numerical simulations of compartmental model (\ref{n.1})-(\ref{n.2}).
(a) Phase trajectories of five uncoupled oscillators are plotted for approximately
one cycle. (b) Timeseries of five uncoupled oscillators. All oscillators were
given identical initial condition lying on the limit cycle of the deterministic
system. Even small noise leads to desynchronization of five oscillators already
after first several cycles. In contrast, simulations of the 
system coupled using nearest-neighbor (c), all-to-all (d), 
and random dissipative (e) coupling
matrices show good coherence. The coherence is better for smaller values
of $\kappa$.
}\label{f.2}
\end{figure}

In the remainder of this section, we present the results of numerical
simulations of (\ref{n.1}) and (\ref{n.2}). We choose the variant 
of the model with partial coupling, i.e., with $\delta=0$.
When $\delta>0$, the model has even better synchronization properties. 
The upper panel of Fig.~\ref{f.2} shows the phase plane and the time
series of five uncoupled oscillators forced by small noise. The initial
condition is chosen on the limit cycle of the deterministic system
and is the same for each oscillator. Fig.~\ref{f.2}a shows phase trajectories
of all five oscillators for approximately one cycle. As one can see
from Fig.~\ref{f.2}b, after a few first cycles, under the influence of
noise the oscillations gradually loose coherence. In contrast,
the trajectories shown in Fig.~\ref{f.2}c and d remain tightly bundled.
In these simulations,  we used nearest-neighbor $\mathbf{D_2}$ and 
all-to-all coupling $\mathbf{D_1}$ respectively. 

To illustrate Theorem~\ref{thm.ql}, for simulations 
shown in Figure~\ref{f.2}e, we use a random dissipative matrix. 
Specifically, we pick the entries of $(-\mathbf{Q})^{1/2}$
from a uniform distribution on $[0,1]$:
$$
(-\mathbf{Q})^{1/2} =
\left(\begin{array}{ccccc}
0.7577 &   0.7060  &  0.8235 &   0.4387 &   0.4898 \\
    0.7431 &   0.0318 &   0.6948 &   0.3816 &   0.4456 \\
    0.3922 &   0.2769 &   0.3171 &   0.7655 &   0.6463 \\
    0.6555 &   0.0462 &   0.9502 &   0.7952 &   0.7094 \\
    0.1712 &   0.0971 &   0.0344 &   0.1869 &   0.7547 
\end{array}
\right)
$$
and let 
\be\lbl{D3}
\mathbf{D_3}=\mathbf{Q\Lambda_0}=
\left(\begin{array}{ccccc}
 -1.0251   & 2.2043 &  -1.6032 &   0.5044   & -0.0804\\
   -0.1264 &   0.2772 &  -0.3006 &   0.2060 &   -0.0562\\
   -1.1549 &   2.5819 &  -1.9613 &   0.5210 &   0.0133 \\
   -0.8807 &   1.9231 &  -1.0823 &   0.0333 &   0.0066 \\
   -0.9049 &   1.8778 &  -1.0060 &   0.3772 &  -0.3441
\end{array}
\right).
\ee
 The trajectories in
Figure~\ref{f.2}e are not as close to each other as in the two previous
plots. 
To see  how the value of $\kappa$ computed
for different network topologies correlates with the 
degree of coherence in our simulations, we compute 
$\kappa (\mathbf{D_1})=0.8$, $\kappa (\mathbf{D_2})=2$,
and $\kappa (\mathbf{D_3})=23.1675$.
In accord with (\ref{v}), the numerics show that coherence
is better for smaller values of $\kappa$.

In conclusion, we relate the class of dissipative matrices
to that of matrices that have been previously known to promote
synchrony. The coupling matrices analyzed in \cite{be, bbh06} are
subject to the constraint that the off diagonal elements are nonnegative.
Note that many of the off diagonal elements of our randomly picked
dissipative matrix $\mathbf{D_3}$ are negative.  A straightforward
albeit tedious calculation shows that
if one chooses a dissipative matrix at random in the way we did in 
this example, the probability that at least one  (or for that matter 
any fixed) off diagonal element is negative, is positive. This shows 
that the class of dissipative matrices is substantially bigger than 
those satisfying sufficient conditions for synchronization
in \cite{be, bbh06}.

\section{Discussion}
\setcounter{equation}{0}
In this Letter, we have identified dissipative operators, a 
class of linear coupling operators  that enforce synchrony in networks 
of oscillators provided that the interactions between oscillators are 
sufficiently strong. Our results apply to a broad class of networks  
including those with asymmetric, time-dependent, and 
nonlinear separable coupling schemes; as well as networks
of local systems with nonperiodic attractors.  Furthermore, we have derived an
analytic estimate (\ref{x}) for the coherence of the network dynamics in the
presence of noise. Robustness to noise is
one of the main indicators gauging physical feasibility of the dynamical 
regimes generated by mathematical models. In this respect, 
(\ref{x}) gives important 
practical information about the factors contributing to the robustness of 
synchronous oscillations to noise. On the other hand, stability of the relevant
 dynamical states is among the key parameters determining the asymptotic 
value of the variance of the trajectories of a randomly perturbed 
dynamical system. For large systems like (\ref{1.2}), analytical estimates 
of the quantities characterizing stability (e.g., Lyapunov exponents) are rare.
By studying the variability of the synchronous regime in a randomly perturbed
problem (\ref{1.2}) and (\ref{sep}), one can infer the degree of stability
of the synchronous solution of the underlying deterministic system.
Specifically, smaller values of $\kappa (\mathbf{D})$ imply better stability
of the synchronous solutions of (\ref{1.2}$)_0$ and (\ref{sep}).
Importantly, $\kappa(\mathbf{D})$ reveals the contribution
of the network topology to the stability of the synchronous state. 
Therefore, using the randomly perturbed model (\ref{1.2}), (\ref{sep}) 
and the main estimate (\ref{x}) may be viewed as a probabilistic method 
for studying stability of the synchronous solutions in the deterministic 
system. 

\vskip 0.5cm
\noindent
{\bf Acknowledgments.} Discussions with Dmitry Kaliuzhnyi-Verbovetskyi
are greatly appreciated. This work was done during sabbatical leave at 
Program of Applied and Computational Mathematics (PACM) at 
Princeton University. The author thanks PACM for hospitality.

\renewcommand{\theequation}{A.\arabic{equation}}
\section*{Appendix. Parameter values for (\ref{n.1}) and (\ref{n.2}) } 
\setcounter{equation}{0}
\label{sec:A}

The equations for the local systems in the neural network (\ref{n.1}) 
and (\ref{n.2}) are adopted from a nondimensional model of a dopamine 
neuron \cite{MC04}.
For biophysical background and details of nondimesionalization, we refer
an interested reader to \cite{MC04}. For the purposes of the present Letter,
the values of several parameters of the original model were modified
to make the oscillations less stiff.  The parameter values used
in the simulations shown in Figure~\ref{f.2} are summarized in  
the following table.

\begin{center}
{\sc Table}
\end{center}
\begin{tabular}{|r|r||r|r||r|r||r|r||r|r||r|r||r|r|}
\hline
$E_1$               & $1$     & 
$E_2$                & $-0.9$  & 
$E_3$                & $-0.3$   & 
$\bar g_1$           & $0.8 $   &              
$\bar g_2$           & $2   $  & 
$\bar g_3$          & $1   $  & 
$g$                  & 0.3    \\
$a_1$                & $-0.35$   & 
$a_2$                &$ 1.4 \cdot 10^{-2}$&                      
$a_3$                &$ 1.8$ & 
$\epsilon$           &$ 0.1$&  
$\tau$               & $5.0 $   & 
$\omega$ &         $5.0$ &
$\sigma$ &        0.0001 \\
\hline
\end{tabular}


\begin{thebibliography}{99}
\bibitem{KS}
I.~Karatzas and S.E.~Shreve, {\it Brownian Motion and Stochastic Calculus},
2nd ed., Springer, New York, 1991.
\bibitem{avr}
V.S.~Afraimovich, N.N.~Verichev, M.I.~Rabinovich, 
Radiophys. Quant. Electron. 29, 795 (1986).
\bibitem{bh84}
J.~Belair and P.~Holmes, Quart. Appl. Math. 42, 193-219 (1983).
\bibitem{ke88}
N.~Kopell, G.B.~Ermentrout, Math. Biosci. 90, 87 (1988).
\bibitem{ku75}
Y.~Kuramoto, Lecture Notes in Physics, vol. 39, edited by
H.~Araki, (Springer, Berlin, 1975) 420--422. 
\bibitem{hale97}
J.K.~Hale, J. Dyn. Diff. Eq. 9, 1 (1997).
\bibitem{pc98}
L.M.~Pecora, T.L.~Caroll, PRL 80, 2109 (1998)
\bibitem{jo}
K.~Josic, Nonlinearity 13, 1321 (2000).
\bibitem{be}
V.N.~Belykh et al., Phys. D 195, 159--187 (2004).
\bibitem{ste}
E.~Steur et al., Phys. D (2009), doi:10.1016/j.physd.2009.08.007. 
\bibitem{pr01}
A.~Pikovsky, M.~Rosenblum, J.~Kurths, Synchronization:
A Universal Concept in Nonlinear sciences, University Press,
Cambridge, 2001.
\bibitem{mmp}
E.~Mosekilde, Yu.~Maistrenko, D.~Postnov, 
Chaotic Synchronization: Applications to Living Systems,
World Scientific Publishing, London, 2002.
\bibitem{izhikevich}
E.M.~Izhikevich E. M., {\it Dynamical Systems in Neuroscience: The 
Geometry of Excitability and Bursting}, Cambridge, Mass: MIT Press, 2007.
\bibitem{coombes}
S.~Coombes, {\it SIAM Journal on Applied Dynamical Systems}, Vol 7, 1101-1129,
2008.
\bibitem{bbh06}
I. Belykh,  V. Belykh, and M. Hasler, 
{\it Physica D}, $\bf 224$, pp. 
42--51 (2006). 
\bibitem{stro}
S.~Strogatz, 
SYNC: The Emerging Science of Spontaneous Order,
Hyperion, New York, 2003.
\bibitem{hale_odes}
J.K.~Hale, Ordinary Differential Equations, 2nd ed.,
Krieger Publishing Company, Malabar, Florida, 1980
\bibitem{HM}
P.~Hitczenko and G.S.~Medvedev, 
{\it SIAM J. Appl. Math.}, 
$\bf 69$(5): 1359-1392, 2009.
\bibitem{M09}
G.S.~Medvedev,  {\it Neural Comp.}, 
$\bf 21$(11): 3057--3078, 2009.
\bibitem{FW}
M.I.~Freidlin and A.D.~Wentzell, 
{\it Random perturbations of dynamical systems}, 
2nd ed., Springer, New York, 1998.
\bibitem{Demidovich}
B.P.~Demidovich, 
{\it Lectures on mathematical theory of stability,}
Nauka, Moscow, 1967. (in Russian)
\bibitem{MC04} G.S.~Medvedev and J.~Cisternas,
{\it Phys. D}, $\bf 194$, 333--356, 2004.
\bibitem{wiggins}
S. Wiggins,{\it  Normally Hyperbolic Manifolds in Dynamical Systems}, 
Springer-Verlag, New York, 1994.
\bibitem{gelfand}
I.M.~Gelfand, {\it Lectures on Linear Algebra},
Interscience Publishers, 1961.
\bibitem{inprep}
G.S.~Medvedev, in preparation. 
\end{thebibliography}
\end{document}